\documentclass{ws-ijmpd-arXiv210505831v6}   

\usepackage{amsmath,amssymb,revsymb,graphicx,dcolumn}

\newcommand{\beq}{\begin{equation}}
\newcommand{\eeq}{\end{equation}}
\newcommand{\beqa}{\begin{eqnarray}}
\newcommand{\eeqa}{\end{eqnarray}}
\newcommand{\bsubeqs}{\begin{subequations}}
\newcommand{\esubeqs}{\end{subequations}}

\def\id{\makebox[0.6ex][l]{$1$}{\rm l}}   

\newcommand\fraclarge[2]{\displaystyle{\frac{\textstyle #1}{\textstyle #2}}}  
\newcommand{\half}{{\textstyle \frac{1}{2}}}    

\begin{document}

\markboth{F.R. Klinkhamer}
{A first look at the bosonic master-field equation of the IIB matrix model}

%
\catchline{}{}{}{}{}
%

\title{A FIRST LOOK AT THE BOSONIC MASTER-FIELD EQUATION\\
       OF THE IIB MATRIX MODEL}

\author{F.R. KLINKHAMER}
\address{Institute for
Theoretical Physics, Karlsruhe Institute of Technology (KIT),\\
76128 Karlsruhe, Germany\\
frans.klinkhamer@kit.edu}

\maketitle


\begin{abstract}
The bosonic large-$N$ master field of the IIB matrix model
can, in principle, give rise to an emergent classical spacetime.
The task is then to calculate this master field as a solution of the
bosonic master-field equation. We consider
a simplified version of the algebraic
bosonic master-field equation
and take dimensionality $D=2$ and matrix size $N=6$.
For an explicit realization of the pseudorandom constants
entering this simplified algebraic equation,
we establish the existence of a solution
and find, after diagonalization of one of the two obtained
matrices, a band-diagonal structure of the other matrix.
\end{abstract}




\section{Introduction}
\label{sec:Intro}

The definitive formulation of nonperturbative superstring
theory, also known
as \mbox{$M$-theory},\cite{Witten1995,HoravaWitten1996}
is still outstanding.
One suggestion is the IKKT matrix model.\cite{IKKT-1997}
That matrix model reproduces the basic structure
of the light-cone string field theory of type-IIB superstrings
and the model is also called the IIB matrix
model.\cite{Aoki-etal-review-1999}
It is, therefore, important to investigate the IIB matrix model.

First results on the partition function of the IIB matrix model
were reported in
Refs.~\refcite{KrauthNicolaiStaudacher1998,AustingWheater2001}
and numerical simulations of the
Lorentzian version of the model were presented in
Refs.~\refcite{KimNishimuraTsuchiya2012,NishimuraTsuchiya2019,%
Hatakeyama-etal2020}
(related numerical results for the Euclidean model
were discussed in Ref.~\refcite{Anagnostopoulos-etal-2020}).
Still, the conceptual question of how classical spacetime
emerges from the IIB matrix model was essentially left unanswered.

We have recently suggested, in the context of the IIB matrix model,
to consider Witten's large-$N$ master
field\cite{Witten1979,Coleman1985,GreensiteHalpern1983,%
Carlson-etal-1983,AlbertyGreensite1984}
as a possible source of the emerging classical
spacetime.\cite{Klinkhamer2020-master}
Assuming the matrices of the bosonic master field to be
known and assuming these matrices to have a band-diagonal
structure, we have shown that, in principle, classical spacetime points
can be extracted and an emerging spacetime metric calculated.
The technical details have been presented in our original
paper,\cite{Klinkhamer2020-master}
further work on the emerging cosmological metric
has been discussed in a follow-up paper,\cite{Klinkhamer2020-reg-bb-IIB-m-m} 
and the heuristics of the spacetime extraction from the master field
has been explained in the recent review.\cite{Klinkhamer2021-APPB-review}

But all these discussions \emph{assume} the
bosonic master field to exist
and to have matrices with a band-diagonal structure
(first hints of a band-diagonal structure were obtained in
numerical results\cite{KimNishimuraTsuchiya2012,NishimuraTsuchiya2019}
for the Lorentzian model).
The goal of our present paper is to embark upon a preliminary
analysis of the
bosonic master-field equation and to check for the
possible appearance of a band-diagonal structure.
As the full bosonic master-field equation from the IIB-matrix-model
is extremely complicated, 
we first consider
a simplified equation and, then, are able to obtain some
nontrivial results. 
Let us, however, emphasize that the present paper is purely exploratory. 
As such, the paper
has a limited scope, which implies, in particular, that all
discussion of statistical issues is postponed to future work.


\section{IIB Matrix Model}
\label{sec:IIB-matrix-model}

\subsection{Action}
\label{subsec:Action}

The IIB matrix model has a \emph{finite number}
of $N \times N$ traceless Hermitian matrices:
ten bosonic matrices $A^{\mu}$ and
essentially eight fermionic (Majorana--Weyl) matrices $\Psi_{\alpha}$.
The partition function $Z$ of the
IIB matrix model is defined by the following ``path''
integral\cite{IKKT-1997,Aoki-etal-review-1999}:%
\bsubeqs\label{eq:IIB-matrix-model-ZwithSgeneral-S-eta}
\beqa
\label{eq:IIB-matrix-model-ZwithSgeneral}
\hspace*{-6mm}
Z &=&\int dA\,d\Psi\;
e^{\displaystyle{-\,S[A,\,\Psi]/ \ell^4}}
= \int dA\;
e^{\displaystyle{-\,S_\text{eff}[A]/ \ell^4}}
\,,
\\[2mm]
\label{eq:IIB-matrix-model-action}
\hspace*{-6mm}
S[A,\,\Psi] &=&
S_\text{bos}[A]+S_\text{ferm}[A,\,\Psi]
\nonumber\\
&=&
-\text{Tr}\,
\Bigg(
\frac{1}{4}\,\big[ A^{\mu} ,\,A^{\nu}    \big]\,
             \big[ A^{\rho},\,A^{\sigma} \big]\,
             \widetilde{\delta}_{\mu\rho}\,\widetilde{\delta}_{\nu\sigma}
+\frac{1}{2}\, \overline{\Psi}_{\beta}\,
\widetilde{\Gamma}^{\mu}_{\beta\alpha}\,\widetilde{\delta}_{\mu\nu}
 \,\big[ A^{\nu},\,\Psi_{\alpha} \big]
\Bigg),
\\[2mm]
\label{eq:IIB-matrix-model-etamunu}
\hspace*{-6mm}
\widetilde{\delta}_{\mu\nu} &=&
\Big[ \text{diag}
\big(  1,\,  1,\, \ldots \,,\,1 \big)
\Big]_{\mu\nu}\,,
\;\;\;\text{for}\;\;\;
\mu, \nu \in \{1,\, 2,\, \ldots\, ,\, 10 \}\,.
\eeqa
\esubeqs
In addition to the partition function $Z$, there are
expectation values of observables,
which will be discussed in Sec.~\ref{subsec:Large-N-bosonic-master-field}.

The fermions appear quadratically
in the action \eqref{eq:IIB-matrix-model-action}.
The fermionic integrals in the first part    
of \eqref{eq:IIB-matrix-model-ZwithSgeneral}  
are then Gaussian 
and can be performed analytically.
In this way, the following effective action is obtained:
\begin{eqnarray}
\label{eq:IIB-matrix-model-Seff}
S_\text{eff}[A]&=&S_\text{bos}[A]+S_\text{ind}[A]\,,
\end{eqnarray}
where the induced term $S_\text{ind}$  may, for example,
contain a high-order term with commutators and anticommutators
of the bosonic matrices.\cite{KrauthNicolaiStaudacher1998}

We have two technical remarks. First,
the model shown in (\ref{eq:IIB-matrix-model-ZwithSgeneral-S-eta})
is the original model with ``Euclidean'' coupling constants
$\widetilde{\delta}_{\mu\nu}$,
but it is also possible to consider a
``Lorentzian'' version\cite{KimNishimuraTsuchiya2012,%
NishimuraTsuchiya2019} with
a complex Feynman phase factor $\exp\left(i\,S/ \ell^4\,\right)$
in the path integral
and coupling constants
$\widetilde{\eta}_{\mu\nu} =
\big[ \text{diag}\big(  -1,\,  1,\, \ldots \,,\,1 \big)
\big]_{\mu\nu}$,
for $\mu, \nu \in \{0,\, 1,\, \ldots\, ,\, 9 \}$, in the action.

Second, a model length scale ``$\ell$'' has been introduced
in (\ref{eq:IIB-matrix-model-ZwithSgeneral-S-eta}), so that
$A^{\mu}$ has the dimension of length and $\Psi_{\alpha}$
the dimension of $(\text{length})^{3/2}$.
From now on, we set
\beq
\label{eq:ell-set-to-1}
\ell=1\,,
\eeq
so that the model contains only dimensionless variables.

Now, the IIB matrix model (\ref{eq:IIB-matrix-model-ZwithSgeneral-S-eta})
just gives numbers,
$Z$ and further expectation values to be discussed later,
while the (dimensionless) matrices $A^{\mu}$ and $\Psi_{\alpha}$
in (\ref{eq:IIB-matrix-model-ZwithSgeneral})
are merely integration variables.
Moreover, there is no obvious small dimensionless parameter
to motivate a saddle-point approximation.
Hence, the following conceptual question arises:
where is the classical spacetime?

For a possible origin of classical spacetime
in the context of the IIB matrix model,
we have suggested\cite{Klinkhamer2020-master}
to revisit an old idea,
the large-$N$ master field of Witten\cite{Witten1979}
(see Ref.~\refcite{Coleman1985} for a review and
Refs.~\refcite{GreensiteHalpern1983,Carlson-etal-1983,AlbertyGreensite1984}
for a selection of subsequent research papers).
In the next two subsections,  
we briefly recall the meaning of this mysterious master
field (a name coined by Coleman\cite{Coleman1985})
and then discuss its ``field'' equation.


\subsection{Large-$N$ bosonic master field}
\label{subsec:Large-N-bosonic-master-field}

Consider the following bosonic observable:
\beq \label{eq:IIB-matrix-model-w-observable}
w^{\mu_{1} \,\ldots\, \mu_{m}} \equiv 
\text{Tr}\,\big( A^{\mu_{1}} \cdots\, A^{\mu_{m}}\big)\,.
\eeq
There are also fermionic
observables (for example, $\text{Tr}\,\overline{\Psi}\,\Psi$),
but here we focus on bosonic observables
of the type  \eqref{eq:IIB-matrix-model-w-observable}.
Then, arbitrary strings of these $w$ observables
have expectation values
\beqa \label{eq:IIB-matrix-model-w-product-vev}
\hspace*{-3mm}&&
\langle
w^{\mu_{1}\,\ldots\,\mu_{m}}\:w^{\nu_{1}\,\ldots\,\nu_{n}}\, \cdots\,
w^{\omega_{1}\,\ldots\,\omega_{z}}
\rangle
=
\frac{1}{Z} \int dA\,
\big(w^{\mu_{1}\,\ldots\,\mu_{m}}\:w^{\nu_{1}\,\ldots\,\nu_{n}}\, \cdots\,
w^{\omega_{1}\,\ldots\,\omega_{z}}\big)\,
e^{\displaystyle{-\,S_\text{eff}}}\,,
\nonumber\\
\hspace*{-3mm}&&
\eeqa
with normalization $\langle\, 1\, \rangle=1$.
For large values of $N$, 
these observables display a remarkable factorization property: 
\beq \label{eq:IIB-matrix-model-w-product-vev-factorized}
\hspace*{-3mm}
\langle
w^{\mu_{1}\,\ldots\,\mu_{m}}\:w^{\nu_{1}\,\ldots\,\nu_{n}}\, \cdots\,
w^{\omega_{1}\,\ldots\,\omega_{z}} \rangle
\;\stackrel{N}{=}\;
\langle w^{\,\mu_{1}\,\ldots\,\mu_{m}}\rangle
\langle w^{\,\nu_{1}\,\ldots\,\nu_{n}}\rangle \, \cdots\,
\langle w^{\,\omega_{1}\,\ldots\,\omega_{z}}\rangle\,,
\eeq
where the equality holds to leading order in $N$.

According to Witten,\cite{Witten1979} the factorization
(\ref{eq:IIB-matrix-model-w-product-vev-factorized}) implies that
the path integrals (\ref{eq:IIB-matrix-model-w-product-vev}) are
saturated by a single configuration,
the so-called master field $\widehat{A}^{\,\mu}$.
To leading order in $N$, the expectation values are then given by
\bsubeqs \label{eq:IIB-matrix-model-w-product-vev-from-master-field}
\beqa
\hspace*{-3mm}
&&\langle
w^{\mu_{1}\,\ldots\,\mu_{m}}\:w^{\nu_{1}\,\ldots\,\nu_{n}}\, \cdots\,
w^{\omega_{1}\,\ldots\,\omega_{z}} \rangle
\;\stackrel{N}{=}\;
\widehat{w}^{\,\mu_{1}\,\ldots\,\mu_{m}}\:
\widehat{w}^{\,\nu_{1}\,\ldots\,\nu_{n}}\, \cdots\,
\widehat{w}^{\,\omega_{1}\,\ldots\,\omega_{z}}\,,
\\[2mm]
&&
\widehat{w}^{\,\mu_{1}\,\ldots\, \mu_{m}}
\equiv
\text{Tr}\,\big( \widehat{A}^{\,\mu_{1}} \cdots\, \widehat{A}^{\,\mu_{m}}\big)\,.
\eeqa
\esubeqs
Hence, we do not have to perform the path integrals
on the right-hand side of (\ref{eq:IIB-matrix-model-w-product-vev}):
we just need ten traceless Hermitian matrices
$\widehat{A}^{\,\mu}$ to get \emph{all} these expectation values
from the simple procedure of replacing each $A^{\,\mu}$
in the observables by the corresponding $\widehat{A}^{\,\mu}$.
Most likely, there is more than one master field,
all these master fields being equivalent
[giving, in the large-$N$ limit,
exactly the same results for all possible observables of the
type \eqref{eq:IIB-matrix-model-w-observable}];
see, e.g., Ref.~\refcite{Carlson-etal-1983} for a discussion
of this point.
But, for definiteness, we will talk, in the following,
only about a single master field.

Now, the meaning of the suggestion
at the end of Sec.~\ref{subsec:Action} is clear:
classical spacetime may reside in the bosonic
master-field matrices $\widehat{A}^{\,\mu}$ of the IIB matrix model.
The heuristics of this idea has been discussed in
Sec.~4.4 of a recent review paper.\cite{Klinkhamer2021-APPB-review}

Next, \emph{assume} that the matrices $\widehat{A}^{\,\mu}$ of the
IIB-matrix-model bosonic master field are known
and that they are approximately band-diagonal.
If, for simplicity, we consider $N=K\,n$ with positive integers 
$K$ and $n$, then it is possible\cite{Klinkhamer2020-master}
to extract from these matrices $\widehat{A}^{\,\mu}$
a discrete set of spacetime points
$\{\widehat{x}^{\,\mu}_{k}\}$ with an index
$k \in \{1,\,\ldots ,\,  K\}$.
These discrete spacetime points sample a smooth
manifold with continuous spacetime coordinates $x^\mu$
and an emergent inverse metric $g^{\mu\nu}(x)$,
for which there is an explicit
expression\cite{Aoki-etal-review-1999,Klinkhamer2020-master}
in terms of the density distribution and correlation functions of
the extracted spacetime points.
The metric $g_{\mu\nu}(x)$ is obtained as matrix inverse of $g^{\mu\nu}(x)$.
The emerging metric may have a Lorentzian signature,
even if the original  matrix model is Euclidean;
see Appendix~B of Ref.~\refcite{Klinkhamer2020-master}
and Appendix~D of Ref.~\refcite{Klinkhamer2021-APPB-review}
for further discussion.

The task is to really \emph{calculate} the
bosonic master-field matrices $\widehat{A}^{\,\mu}$ of the
IIB matrix model and, if possible, to establish
a band-diagonal structure. For this calculation, we
need the  ``field'' equation for these master matrices.


\subsection{Bosonic master-field equation}
\label{subsec:Bosonic master-field-equation}

Building on previous work by
Greensite and Halpern,\cite{GreensiteHalpern1983}
we have obtained the IIB-matrix-model
bosonic master field
in the following ``quenched'' form\cite{Klinkhamer2020-master}:
\bsubeqs\label{eq:IIB-matrix-model-master-field-algebraic-equation}
\beq
\label{eq:IIB-matrix-model-master-field}
\widehat{A}^{\;\rho}_{\;kl}
=
e^{\displaystyle{i\,(\widehat{p}_{k}-\widehat{p}_{l})\tau_\text{eq}}}
\;\,
\widehat{a}^{\;\rho}_{\;kl}\,,
\eeq
where the matrix indices $k$ and $l$
take values from $\{  1,\, \, \ldots \, N \}$
and the directional index $\rho$ runs
over $\{  1,\,  2,\, \ldots \,,\,D \}$.
The dimensionless time
$\tau_\text{eq}$ in \eqref{eq:IIB-matrix-model-master-field}
must have a sufficiently large value in order to
represent an equilibrium situation
($\tau$ is the fictitious Langevin
time of the stochastic-quantization procedure).
The $\tau$-independent matrix $\widehat{a}^{\;\rho}$
on the right-hand side of \eqref{eq:IIB-matrix-model-master-field}
solves the following algebraic equation\cite{Klinkhamer2020-master}:
\begin{eqnarray}
\label{eq:IIB-matrix-model-algebraic-equation}
i\,\big(\widehat{p}_{k}-\widehat{p}_{l}\big)\;
\widehat{a}^{\;\rho}_{\;kl}
&=&
-\left.\frac{\delta S_\text{eff}}{\delta A_{\rho\;lk}}
 \right|_{A=\widehat{a}}\;
+\widehat{\eta}^{\;\rho}_{\;kl}\,,
\end{eqnarray}
\esubeqs
in terms of the master momenta
$\widehat{p}_{k}$ (uniform random numbers)
and the 
master-noise   
matrices $\widehat{\eta}^{\;\rho}_{\;kl}$
(Gaussian random numbers);
further details and references
can be found in Ref.~\refcite{GreensiteHalpern1983}.

The matrices $\widehat{a}^{\;\rho}$ are
$N \times N$ traceless Hermitian matrices
and the number of real bosonic degrees of freedom is
\beq\label{eq:Ndof}
N_\text{d.o.f.} = D\,\big(N^2-1\big)\,.
\eeq
These degrees of freedom are determined by  
the algebraic equation
\eqref{eq:IIB-matrix-model-algebraic-equation} for fixed random
constants $\widehat{p}_{k}$ and $\widehat{\eta}^{\;\rho}_{\;kl}$.
It remains to solve this algebraic equation, which is not quite trivial,
as there is a complicated high-order term in $S_\text{eff}$ 
from fermion induction effects.  
In this paper, we will take a first step
by considering a simplified version
of \eqref{eq:IIB-matrix-model-algebraic-equation}.


\section{Simplified Algebraic Equation}
\label{sec:Simplified-algebraic-equation}

\subsection{General case}
\label{subsec:General-case}

For $N \times N$ traceless Hermitian matrices $\widehat{a}^{\;\rho}$
with index $\rho$ running over $\{  1,\,  2,\, \ldots \,,\,D \}$,
we will consider the following simplified algebraic equation:
\begin{subequations}
\label{eq:simplified-equation-gtildemunu-stilde-Eucl}
\begin{eqnarray}
\label{eq:simplified-equation}
\hspace*{-5mm}
i\,\big(\widehat{p}_{k}-\widehat{p}_{l}\big)\;
\widehat{a}^{\;\rho}_{\;kl}
&=&
\widetilde{g}_{\mu\nu}\,
\Big[\widehat{a}^{\,\mu},\,\big[\widehat{a}^{\,\nu},\,
\widehat{a}^{\;\rho}\big]\Big]_{kl}
+\widehat{\eta}^{\;\rho}_{\;kl}\,,
\\[2mm]
\label{eq:simplified-equation-gtildemunu}
\hspace*{-5mm}
\widetilde{g}_{\mu\nu} &=&
\Big[ \text{diag}\left(\widetilde{g}_{11},\,
\widetilde{g}_{22},\, \ldots \,,\,\widetilde{g}_{DD} \right)\Big]_{\mu\nu}
=
\Big[ \text{diag}\left(\widetilde{s},\,
1,\, \ldots \,,\,1 \right)\Big]_{\mu\nu}\,,
\\[2mm]
\label{eq:simplified-equation-stilde-Eucl}
\hspace*{-5mm}
\widetilde{s} &=&  1\,,
\end{eqnarray}
\end{subequations}
where $k$ and $l$ are matrix indices running over
$\{  1,\,  \ldots \,,\,N \}$ and where
we omit matrix indices inside the double commutator
on the right-hand side of \eqref{eq:simplified-equation}.
The choice $\widetilde{s}=1$ corresponds to ``Euclidean''
coupling constants $\widetilde{g}_{\mu\nu}$
from \eqref{eq:simplified-equation-gtildemunu}.
The pseudorandom numbers $\widehat{p}_{k}$
and $\widehat{\eta}^{\;\rho}_{\;kl}$
will be specified in Sec.~\ref{subsec:Special-case-D2-N4}
for a special case which can be easily generalized.

The  crucial simplification
of \eqref{eq:simplified-equation-gtildemunu-stilde-Eucl},
compared to the full algebraic
equation \eqref{eq:IIB-matrix-model-algebraic-equation},
is that the effects of the fermions are neglected, which are
contained in the $S_\text{ind}$ contribution to the
effective action \eqref{eq:IIB-matrix-model-Seff}.
Remark also that our simplified algebraic equation
\eqref{eq:simplified-equation-gtildemunu-stilde-Eucl}
resembles the ``classical equation'' studied in
Ref.~\refcite{Hatakeyama-etal2020},
as given by Eq.~(2.6) in that reference
with implicit ``Lorentzian'' coupling constants.


\subsection{Special case: $D=2$ and $N=4$}
\label{subsec:Special-case-D2-N4}

In order to be specific, let us first consider the case
\beq\label{eq:parameters-D2-N4}
\big\{ D,\,  N\big\}
=
\big\{ 2,\,  4\big\}\,.
\eeq
The discussion of this subsection trivially extends to  
larger values of $D$ and $N$.
For the case \eqref{eq:parameters-D2-N4}, we parameterize the matrices
$\widehat{a}^{\,1}$ and $\widehat{a}^{\,2}$ as follows:
\bsubeqs\label{eq:ahatmu-Ansaetze-N4}
\beqa
\widehat{a}^{\,1}
&=&
\left(
\renewcommand{\arraycolsep}{0.5pc} 
\renewcommand{\arraystretch}{1.0}  
  \begin{array}{cccc}
a_{11} & a_{12}+i\,A_{12} & a_{13}+i\,A_{13} & a_{14} + i \,A_{14}\\
a_{12} - i \,A_{12} & a_{22} & a_{23} + i \,A_{23} & a_{24} + i \,A_{24}\\
a_{13} - i \,A_{13} & a_{23} - i \,A_{23} & a_{33} & a_{34} + i \,A_{34}\\
a_{14} - i \,A_{14} & a_{24} - i \,A_{24} & a_{34} - i \,A_{34} & -a_{11} - a_{22} - a_{33}\\
  \end{array}
\right)\,,
\\[2mm]
\widehat{a}^{\,2} &=&
\left(
\renewcommand{\arraycolsep}{0.5pc} 
\renewcommand{\arraystretch}{1.0}  
  \begin{array}{cccc}
b_{11} & b_{12}+i\,B_{12} & b_{13}+i\,B_{13} & b_{14} + i \,B_{14}\\
b_{12} - i \,B_{12} & b_{22} & b_{23} + i \,B_{23} & b_{24} + i \,B_{24}\\
b_{13} - i \,B_{13} & b_{23} - i \,B_{23} & b_{33} & b_{34} + i \,B_{34}\\
b_{14} - i \,B_{14} & b_{24} - i \,B_{24} & b_{34} - i \,B_{34} & -b_{11} - b_{22} - b_{33}\\
  \end{array}
\right)\,,
\eeqa
\esubeqs
in terms of the 15 real variables
$\{  a_{11},\, a_{12},\, a_{22},\, \ldots \,   ,\,A_{34} \}$
and the 15 real variables
$\{  b_{11},$ $b_{12}$, $b_{22},\, \ldots \,   ,\,B_{34} \}$.
Hence, the number of unknowns is $30$.

Similarly, the master coupling constants are parameterized as follows:
\bsubeqs\label{eq:phat-etahatmu-Ansaetze-N4}
\beqa
\widehat{p}
&=&
\left\{ \widehat{p}_{1} ,\,\widehat{p}_{2}  ,\, \widehat{p}_{3}  ,\, \widehat{p}_{4}   \right\}\,,
\\[2mm]
\widehat{\eta}^{\,1}
&=&
\left(
\renewcommand{\arraycolsep}{0.5pc} 
\renewcommand{\arraystretch}{1.0}  
  \begin{array}{cccc}
e_{11} & e_{12}+i\,E_{12} & e_{13}+i\,E_{13} & e_{14} + i \,E_{14}\\
e_{12} - i \,E_{12} & e_{22} & e_{23} + i \,E_{23} & e_{24} + i \,E_{24}\\
e_{13} - i \,E_{13} & e_{23} - i \,E_{23} & e_{33} & e_{34} + i \,E_{34}\\
e_{14} - i \,E_{14} & e_{24} - i \,E_{24} & e_{34} - i \,E_{34} & -e_{11} - e_{22} - e_{33}\\
  \end{array}
\right)\,,
\\[2mm]
\widehat{\eta}^{\,2}
&=&
\left(
\renewcommand{\arraycolsep}{0.5pc} 
\renewcommand{\arraystretch}{1.0}  
  \begin{array}{cccc}
f_{11} & f_{12}+i\,F_{12} & f_{13}+i\,F_{13} & f_{14} + i \,F_{14}\\
f_{12} - i \,F_{12} & f_{22} & f_{23} + i \,F_{23} & f_{24} + i \,F_{24}\\
f_{13} - i \,F_{13} & f_{23} - i \,F_{23} & f_{33} & f_{34} + i \,F_{34}\\
f_{14} - i \,F_{14} & f_{24} - i \,F_{24} & f_{34} - i \,F_{34} & -f_{11} - f_{22} - f_{33}\\
  \end{array}
\right)\,.
\eeqa
\esubeqs
The entries in \eqref{eq:phat-etahatmu-Ansaetze-N4}
are given by pseudorandom rational numbers
with ranges $[-1/2,\,1/2]$
for the master momenta and $[-1,\,1]$ for the master noise:  
\bsubeqs\label{eq:phat-etahat-random-numbers}
\beqa\label{eq:phat-etahat-random-numbers-phat}
\widehat{p}_{k}
&=&
\left(\frac{\text{randominteger}[-500,\,+500]}{1000}\right)_{k} \,,
\\[2mm]
\label{eq:phat-etahat-random-numbers-ekl}
e_{kl}
&=&
\left(\frac{\text{randominteger}[-1000,\,+1000]}{1000}\right)_{kl}\,,
\\[2mm]
\label{eq:phat-etahat-random-numbers-Ekl}
E_{kl}
&=&
\left(\frac{\text{randominteger}[-1000,\,+1000]}{1000}\right)_{kl}\,,
\eeqa
\beqa
\label{eq:phat-etahat-random-numbers-fkl}
f_{kl}
&=&
\left(\frac{\text{randominteger}[-1000,\,+1000]}{1000}\right)_{kl}\,,
\\[2mm]
\label{eq:phat-etahat-random-numbers-Fkl}
F_{kl}
&=&
\left(\frac{\text{randominteger}[-1000,\,+1000]}{1000}\right)_{kl}\,,
\eeqa
\esubeqs
where the pseudorandom integers on the right-hand sides
are taken from uniform distributions with ranges as indicated.
The practical reason for taking rational numbers
is that we can then easily write
down their exact values, whereas real numbers would require
an infinite number of digits (or an implicit defining relation,
as for the irrational number $\sqrt{2}$\,).

Strictly speaking, the pseudorandom master-noise numbers
$e_{kl}$, $E_{kl}$, $f_{kl}$, and $F_{kl}$
must be taken from a Gaussian (normal) distribution,
but here we have used, for simplicity, a uniform distribution
with a finite range.
For later reference, we can mention that an improved procedure
would use a truncated Gaussian distribution,
\beq
\label{eq:truncated-Gaussian-distribution}
P_\text{trunc-Gauss}(x)=
\begin{cases}
\nu\;e^{\displaystyle{-\half\;x^2/\sigma^2}} \,,
&  \;\;\text{for}\;\; |x| \leq x_\text{trunc} \,,
 \\[2mm]
0 \,,
&  \;\;\text{for}\;\; |x| > x_\text{trunc} \,,
\end{cases}
\eeq
with spread $\sigma>0$, cut-off value $x_\text{trunc}>0$,
and normalization factor $\nu=\nu(\sigma,\, x_\text{trunc})$.
In principle, we should have $x_\text{trunc}\gg \sigma$,
so that $\sigma$ approaches the standard deviation.
As we are primarily interested in establishing the
existence of a nontrivial solution and giving a qualitative
discussion of its properties, we have used a simple
and explicit procedure
in \eqref{eq:phat-etahat-random-numbers-ekl}%
--\eqref{eq:phat-etahat-random-numbers-Fkl}
with rational numbers $n/1000$, for integer $n \in [-1000,\,+1000]$,
taken from a uniform distribution [which corresponds to
$x_\text{trunc}=1$ and $\sigma \gg 1$
in \eqref{eq:truncated-Gaussian-distribution}].
An in-depth discussion of the
statistical aspects of the obtained
solutions requires significantly larger values of
$N$ and also a larger dimensionality, $D=10$.

From \eqref{eq:simplified-equation-gtildemunu-stilde-Eucl}
and \eqref{eq:parameters-D2-N4}, we have 30 coupled
algebraic equations for the 30 real unknowns
$\{  a_{11},\,  \ldots \,$,  $B_{34} \}$.
It appears impossible to obtain a general analytic solution
in terms of the 34 constants
$\{  \widehat{p}_{k},\,  \ldots \,,\,F_{34} \}$.
(Remark that an exact solution was obtained in
Sec.~5 of Ref.~\refcite{GreensiteHalpern1983}
for a single-matrix model at $N=2$.)
In our case with two matrices and $N=4$,
we will consider the 30 coupled algebraic equations
with an explicit choice for the 34 pseudorandom constants.
The reader who is allergic to seeing too many numbers
may skip ahead to Sec.~\ref{sec:Conclusion}.


\section{Numerical Solutions}
\label{sec:Numerical solutions}

\subsection{$D=2$ and $N=4$}
\label{subsec:D2-and-N4}


We present, here, a representative numerical solution
of the 30 coupled algebraic equations
mentioned at the end of Sec.~\ref{subsec:Special-case-D2-N4}.
Take, for example, the following pseudorandom constants:
\bsubeqs\label{eq:pseudorandom-constants-N4}
\beqa
\hspace*{-12mm}&& \widehat{p}_\text{num}
=
\left\{ \frac{159}{1000},\; -\frac{73}{250},\;- \frac{141}{500},\;
- \frac{209}{500}  \right\}\,,
\eeqa
\beqa
\hspace*{-12mm}&&\widehat{\eta}^{\,1}_\text{num}
=
\left(
\renewcommand{\arraycolsep}{0.15pc} 
\renewcommand{\arraystretch}{2.0}  
  \begin{array}{cccc}
-\fraclarge{97}{200} &
\fraclarge{209}{500} - \fraclarge{111\,i}{1000}&
\fraclarge{229}{1000} + \fraclarge{221\,i}{500}&
- \fraclarge{23}{50}   - \fraclarge{23\,i}{40}
\\
\fraclarge{209}{500} + \fraclarge{111\,i}{1000}&
- \fraclarge{1}{125} &
- \fraclarge{923}{1000}   + \fraclarge{333\,i}{500}&
- \fraclarge{77}{500}   + \fraclarge{169\,i}{500}
\\
\fraclarge{229}{1000} - \fraclarge{221\,i}{500}&
- \fraclarge{923}{1000}   - \fraclarge{333\,i}{500}&
- \fraclarge{273}{500} &
- \fraclarge{471}{500}   + \fraclarge{681\,i}{1000}
\\
- \fraclarge{23}{50}   + \fraclarge{23\,i}{40}&
- \fraclarge{77}{500}   - \fraclarge{169\,i}{500}&
- \fraclarge{471}{500}   - \fraclarge{681\,i}{1000}&
 \fraclarge{1039}{1000}\\
  \end{array}
\right)\,,
\eeqa
\beqa
\hspace*{-12mm}&&\widehat{\eta}^{\,2}_\text{num}
=
\left(
\renewcommand{\arraycolsep}{0.15pc} 
\renewcommand{\arraystretch}{2.0}  
  \begin{array}{cccc}
 - \fraclarge{701}{1000} &
   \fraclarge{543}{1000} + \fraclarge{463\,i}{500}&
   - \fraclarge{419}{1000}   - \fraclarge{453\,i}{1000}&
   \fraclarge{307}{1000} + \fraclarge{339\,i}{1000}
\\
\fraclarge{543}{1000} - \fraclarge{463\,i}{500}&
- \fraclarge{1}{2} &
- \fraclarge{559}{1000}   + \fraclarge{299\,i}{1000}&
 \fraclarge{249}{250} - \fraclarge{301\,i}{1000}
\\
- \fraclarge{419}{1000}   + \fraclarge{453\,i}{1000}&
- \fraclarge{559}{1000}   - \fraclarge{299\,i}{1000}&
\fraclarge{191}{250}&
- \fraclarge{171}{200}   + \fraclarge{439\,i}{1000}
\\
\fraclarge{307}{1000} - \fraclarge{339\,i}{1000}&
   \fraclarge{249}{250} + \fraclarge{301\,i}{1000}&
   - \fraclarge{171}{200}   - \fraclarge{439\,i}{1000}&
   \fraclarge{437}{1000}
\\
 \end{array}
\right).
\eeqa
\esubeqs
Then, a numerical solution of the simplified algebraic equation
\eqref{eq:simplified-equation-gtildemunu-stilde-Eucl},
for the parameters \eqref{eq:parameters-D2-N4},  
is given by
\bsubeqs\label{eq:ahat1numsol-ahat2numsol-N4}
\beqa
\hspace*{-12mm}&&
\widehat{a}^{\,1}_\text{num-sol}
=
\nonumber\\[1mm]
\hspace*{-12mm}&&
\left(
\renewcommand{\arraycolsep}{0.25pc} 
\renewcommand{\arraystretch}{1.0}  
  \begin{array}{cccc}
 -0.674 & -0.024-0.432 \, i & 0.763+0.788 \, i & 0.071+0.578 \, i \\
 -0.024+0.432 \, i & 0.382 & 0.791+0.512 \, i & 0.67-1.57 \, i \\
 0.763-0.788 \, i & 0.791-0.512 \, i & 0.565 & 0.566+0.698 \, i \\
 0.071-0.578 \, i & 0.67+1.57 \, i & 0.566-0.698 \, i & -0.274 \\
 \end{array}
\right),
\eeqa
\beqa
\hspace*{-12mm}&&\widehat{a}^{\,2}_\text{num-sol}
=
\nonumber\\[1mm]
\hspace*{-12mm}&&
\left(
\renewcommand{\arraycolsep}{0.25pc} 
\renewcommand{\arraystretch}{1.0}  
  \begin{array}{cccc}
 0.690 & 0.128-0.292 \, i & 0.217+0.356 \, i & 0.536-0.437 \, i \\
 0.128+0.292 \, i & -0.652 & 0.593+0.518 \, i & -0.314+0.776 \, i \\
 0.217-0.356 \, i & 0.593-0.518 \, i & 0.631 & -0.041+0.201 \, i \\
 0.536+0.437 \, i & -0.314-0.776 \, i & -0.041-0.201 \, i & -0.669 \\
  \end{array}
\right)\,,
\eeqa
\esubeqs
where only two or three  
significant digits are shown
(typically, we have a 24-digit working precision).
Incidentally, the solution \eqref{eq:ahat1numsol-ahat2numsol-N4}
is not unique, as there is, at least, one other solution.

Considering the absolute values of the entries in the
matrices \eqref{eq:ahat1numsol-ahat2numsol-N4}, we have
\bsubeqs\label{eq:ABSahatmunumsol-N4}
\beqa\label{eq:ABSahat1numsol-N4}
\hspace*{-5mm}&&\text{Abs}\left[\widehat{a}^{\,1}_\text{num-sol}\right]
=
\left(
\renewcommand{\arraycolsep}{0.5pc} 
\renewcommand{\arraystretch}{1.0}  
  \begin{array}{cccc}
 0.674 & 0.433 & 1.10 & 0.582 \\
 0.433 & 0.382 & 0.942 & 1.71 \\
 1.10 & 0.942 & 0.565 & 0.899 \\
 0.582 & 1.71 & 0.899 & 0.274 \\
  \end{array}
\right)\,,
\\[2mm]
\label{eq:ABSahat2numsol-N4}
\hspace*{-5mm}&&\text{Abs}\left[\widehat{a}^{\,2}_\text{num-sol}\right]
=
\left(
\renewcommand{\arraycolsep}{0.5pc} 
\renewcommand{\arraystretch}{1.0}  
  \begin{array}{cccc}
 0.690 & 0.319 & 0.417 & 0.691 \\
 0.319 & 0.652 & 0.788 & 0.837 \\
 0.417 & 0.788 & 0.631 & 0.206 \\
 0.691 & 0.837 & 0.206 & 0.669 \\
  \end{array}
\right)\,,
\eeqa
\esubeqs
and we see no obvious band-diagonal structure.
(Recall that the diagonal is singled out by the
Hermiticity condition on the matrix $\widehat{a}^{\,1}$,
making the entries on its diagonal real, and similarly
for the matrix $\widehat{a}^{\,2}$.)
The presence or absence of a band-diagonal structure
can be quantified by the following averages and ratios:
\bsubeqs\label{eq:ABSamu-averages-N4}
\beqa
\hspace*{-11mm}&&
\Big\{
\langle \text{all} \rangle,\,
\langle \text{band-diag} \rangle,\,
\langle \text{off-band-diag} \rangle,\,
\text{ratio}
\Big\}_{\text{Abs}\left[\widehat{a}^{\,1}_\text{num-sol}\right]}^{(N=4)}
=
\nonumber\\[1mm]
\hspace*{-12mm}&&
\left\{ 0.827,\, 0.644,\, 1.13,\, 0.570 \right\},
\\[2mm]
\hspace*{-11mm}&&
\Big\{
\langle \text{all} \rangle,\,
\langle \text{band-diag} \rangle,\,
\langle \text{off-band-diag} \rangle,\,
\text{ratio}
\Big\}_{\text{Abs}\left[\widehat{a}^{\,2}_\text{num-sol}\right]}^{(N=4)}
=
\nonumber\\[1mm]
\hspace*{-11mm}&&
\left\{ 0.572,\, 0.527,\, 0.649,\, 0.812 \right\},
\eeqa
\esubeqs
which give, respectively,
the average absolute value of all (16) matrix entries,
the average absolute value of the band-diagonal (3+4+3) matrix entries,
the average absolute value of the off-band-diagonal (3+3) matrix entries,
and the ratio of the band-diagonal value over the off-band-diagonal value.
The ratios shown as the last entries of \eqref{eq:ABSamu-averages-N4}
are of order unity.

Next, we diagonalize one of the matrices, while ordering the eigenvalues,
and look at the other matrix
to see if it has a band-diagonal structure (even for the very small value
of $N$ we are considering). Recall that the diagonalization is
achieved by use of a similarity transformation, 
\beq
\label{eq:ahatrho-similarity-transformed}
\widehat{a}^{\,\rho\;\text{(new)}}
= S \cdot \widehat{a}^{\;\rho} \cdot S^{-1}\,,
\eeq
where an appropriate choice of the matrix $S$ can make one
of the matrices diagonal (see, e.g., Sec.~11.0 of
Ref.~\refcite{Press-etal-1992} for a clear discussion).

If we diagonalize and order $\widehat{a}^{\,1}_\text{num-sol}$
(the new matrices are denoted by a prime), we get
\bsubeqs\label{eq:amuprimenumsol-N4}
\beqa
\hspace*{-3mm}&&\widehat{a}^{\;\prime\,1}_\text{num-sol}
= S_{1} \cdot \widehat{a}^{\,1}_\text{num-sol} \cdot S_{1}^{-1}
=
\text{diag}\big( -2.40 ,\, -1.19  ,\, 1.60  ,\,  1.99  \big)\,,
\\[2mm]
\label{eq:a2primenumsol-N4}
\hspace*{-5mm}&&\widehat{a}^{\;\prime\,2}_\text{num-sol}
= S_{1} \cdot \widehat{a}^{\,2}_\text{num-sol} \cdot S_{1}^{-1}=
\nonumber\\[1mm]
\hspace*{-3mm}&&
\left(   
\renewcommand{\arraycolsep}{0.15pc} 
\renewcommand{\arraystretch}{1.0}  
  \begin{array}{cccc}
 -0.507 & 0.249+0.415 \, i & 0.0355-0.0451 \, i & -0.0316+0.0235 \, i \\
 0.249-0.415 \, i & 0.847 & 0.181+0.031 \, i & -0.0756-0.0105 \, i \\
 0.0355+0.0451 \, i & 0.181-0.031 \, i & 0.432 & 1.330+0.258 \, i \\
 -0.0316-0.0235 \, i & -0.0756+0.0105 \, i & 1.330-0.258 \, i & -0.772 \\
  \end{array}
\right),
\nonumber\\
\hspace*{-3mm}&&
\eeqa
\esubeqs
where $S_{1}$ is a short-hand notation for $S_{1,\,\text{num}}$ and
where, again, only 
two or three  
significant digits are shown.

For the record, we give the absolute values of the
entries of the matrix \eqref{eq:a2primenumsol-N4},
\beq\label{eq:Absa2primenumsol-N4}
\hspace*{-0mm}
\text{Abs}\left[\widehat{a}^{\;\prime\,2}_\text{num-sol}\right]
=
\left(
\renewcommand{\arraycolsep}{0.5pc} 
\renewcommand{\arraystretch}{1.0}  
  \begin{array}{cccc}
 0.507 & 0.484 & 0.0574 & 0.0394 \\
 0.484 & 0.847 & 0.183 & 0.0763 \\
 0.0574 & 0.183 & 0.432 & 1.35 \\
 0.0394 & 0.0763 & 1.35 & 0.772 \\
  \end{array}
\right)\,,
\eeq
where we see that the far-off-diagonal elements
(e.g., at positions $13$, $14$, and $24$) are
rather small in comparison to those close to the diagonal,
which is not the case for the original matrix \eqref{eq:ABSahat2numsol-N4}.
Again, this can be quantified by the following averages and ratio:
\beqa
\label{eq:a1-diag-averages-N4}
\hspace*{-3mm}&&
\Big\{
\langle \text{all} \rangle,\,
\langle \text{band-diag} \rangle,\,
\langle \text{off-band-diag} \rangle,\,
\text{ratio}
\Big\}_{\text{Abs}\left[\widehat{a}^{\;\prime\,2}_\text{num-sol}\right]}^{(N=4)}
=
\nonumber\\
\hspace*{-3mm}&&
\left\{ 0.434,\, 0.660,\, 0.0577,\, 11.4 \right\}\,,
\eeqa
where the meaning of the quantities has been explained
on the lines below \eqref{eq:ABSamu-averages-N4}.
The ratio of the band-diagonal value over the off-band-diagonal value
from \eqref{eq:a1-diag-averages-N4} is of order $10$.

Similar results are obtained if the other matrix
$\widehat{a}^{\,2}_\text{num-sol}$ is diagonalized and
ordered. The new matrices are denoted by a double prime
and are obtained by a similarity transformation
\eqref{eq:ahatrho-similarity-transformed}
with an appropriate  matrix $S_{2}$.
The corresponding averages and ratio are
\beqa
\label{eq:a2-diag-averages-N4}
\hspace*{-3mm}&&
\Big\{
\langle \text{all} \rangle,\,
\langle \text{band-diag} \rangle,\,
\langle \text{off-band-diag} \rangle,\,
\text{ratio}
\Big\}_{\text{Abs}\left[\widehat{a}^{\;\prime\prime\,1}_\text{num-sol}\right]}^{(N=4)}
=
\nonumber\\
\hspace*{-3mm}&&
\left\{ 0.655,\, 0.977,\, 0.120,\, 8.17 \right\}\,,
\eeqa
where the ratio of the band-diagonal value over the off-band-diagonal value
is of order $10$.

For both choices of the bases [matrices $S=S_{1}$ or $S=S_{2}$
in the similarity transformation \eqref{eq:ahatrho-similarity-transformed}],
we see a more or less band-diagonal structure
if one of the matrices is diagonalized.
With different realizations of the pseudorandom constants,
there is significant scatter in the values of the ratios mentioned
in \eqref{eq:a1-diag-averages-N4} and \eqref{eq:a2-diag-averages-N4},
but all values are apparently above unity.
As mentioned before, a proper analysis of statistical issues
must wait for solutions at larger values of $N$ and $D$.

These results were obtained
from the simplified algebraic equation \eqref{eq:simplified-equation}
with ``Euclidean'' coupling constants $\widetilde{g}_{\mu\nu}$
from \eqref{eq:simplified-equation-gtildemunu}
having $\widetilde{s} =1$.  Similar results are obtained with
``Lorentzian'' coupling constants  $\widetilde{g}_{\mu\nu}$
 having $\widetilde{s} =- 1$.


\subsection{$D=2$ and $N=6$}
\label{subsec:D2-and-N6}


The numerical solution of Sec.~\ref{subsec:D2-and-N4}
has suggested that a band-diagonal structure appears
if one of the matrices $\widehat{a}^{\;\rho}$ is diagonalized.
But the matrix size considered ($N=4$) was rather small,
with the number of off-band-diagonal elements
($N_\text{off-band-diag}=6$) being smaller than
the number of band-diagonal elements ($N_\text{band-diag}=10$).
In principle, we would like to have
$N_\text{off-band-diag}=\text{O}(N^2)  \gg
N_\text{band-diag}=\text{O}(N)$, 
for very large $N$ and fixed width $\Delta N$ of the band diagonal. 
We take a modest step in the right direction by
using $N=6$ and $N_\text{off-band-diag}=20 > N_\text{band-diag}=16$.

We present a representative numerical solution
for the case
\beq\label{eq:parameters-D2-N6}
\big\{ D,\,  N\big\}
=
\big\{ 2,\,  6 \big\}
\eeq
and a particular choice of pseudorandom constants
$\{ \widehat{p}_\text{num},\,\widehat{\eta}^{\,1}_\text{num},\,
\widehat{\eta}^{\,2}_\text{num}\}$, which are
shown explicitly in \ref{app:D2-and-N6}.
The simplified algebraic
equation \eqref{eq:simplified-equation-gtildemunu-stilde-Eucl}
with parameters \eqref{eq:parameters-D2-N6} then gives  
a numerical solution $\widehat{a}^{\,1}_\text{num-sol}$
and $\widehat{a}^{\,2}_\text{num-sol}$, which is
also shown explicitly in \ref{app:D2-and-N6}.

Here, we only display the absolute values of the entries
of these last matrices,
\bsubeqs\label{eq:ABSahat1numsol-ABSahat2numsol-N6}
\beqa\label{eq:ABSahat1numsol-N6}
\hspace*{-5mm}&&
\text{Abs}\left[\widehat{a}^{\,1}_\text{num-sol}\right]
=
\left(
\renewcommand{\arraycolsep}{0.5pc} 
\renewcommand{\arraystretch}{1.0}  
\begin{array}{cccccc}
 0.234 & 0.416 & 0.636 & 1.12 & 0.831 & 0.232 \\
 0.416 & 0.721 & 0.949 & 0.566 & 0.611 & 0.366 \\
 0.636 & 0.949 & 0.445 & 0.754 & 1.20 & 0.623 \\
 1.12 & 0.566 & 0.754 & 0.0504 & 0.114 & 0.620 \\
 0.831 & 0.611 & 1.20 & 0.114 & 0.256 & 0.631 \\
 0.232 & 0.366 & 0.623 & 0.620 & 0.631 & 0.248 \\
\end{array}
\right)\,,
\eeqa
\beqa\label{eq:ABSahat2numsol-N6}
\hspace*{-5mm}&&
\text{Abs}\left[\widehat{a}^{\,2}_\text{num-sol}\right]
=
\left(
\renewcommand{\arraycolsep}{0.5pc} 
\renewcommand{\arraystretch}{1.0}  
\begin{array}{cccccc}
 0.560 & 0.926 & 0.560 & 0.400 & 0.565 & 0.620 \\
 0.926 & 0.866 & 0.701 & 0.813 & 0.624 & 1.11 \\
 0.560 & 0.701 & 0.623 & 1.04 & 0.0857 & 0.624 \\
 0.400 & 0.813 & 1.04 & 0.459 & 0.469 & 0.911 \\
 0.565 & 0.624 & 0.0857 & 0.469 & 0.400 & 0.485 \\
 0.620 & 1.11 & 0.624 & 0.911 & 0.485 & 1.19 \\
\end{array}
\right)\,,
\eeqa
\esubeqs
where we do not observe any obvious band-diagonal structure.
In fact, from the values given in
\eqref{eq:ABSahat1numsol-ABSahat2numsol-N6},
we calculate the following averages and ratios:
\bsubeqs\label{eq:ABSamu-averages-N6}
\beqa
\hspace*{-11mm}&&
\Big\{
\langle \text{all} \rangle,\,
\langle \text{band-diag} \rangle,\,
\langle \text{off-band-diag} \rangle,\,
\text{ratio}
\Big\}_{\!\text{Abs}\left[\widehat{a}^{\,1}_\text{num-sol}\right]}^{\!(N=6)}
=
\nonumber\\[1mm]
\hspace*{-3mm}&&
\left\{ 0.592,\, 0.480,\, 0.681,\, 0.705 \right\},
\\[2mm]
\hspace*{-3mm}&&
\Big\{
\langle \text{all} \rangle,\,
\langle \text{band-diag} \rangle,\,
\langle \text{off-band-diag} \rangle,\,
\text{ratio}
\Big\}_{\!\text{Abs}\left[\widehat{a}^{\,2}_\text{num-sol}\right]}^{\!(N=6)}
=
\nonumber\\[1mm]
\hspace*{-3mm}&&
\left\{ 0.666,\, 0.709,\, 0.631,\, 1.12 \right\},
\eeqa
\esubeqs
which give, respectively,
the average absolute value of all (36) matrix entries,
the average absolute value of the band-diagonal (5+6+5) matrix entries,
the average absolute value of the off-band-diagonal (10+10) matrix entries,
and the ratio of the band-diagonal value over the off-band-diagonal value.
The ratios shown as the last entries of \eqref{eq:ABSamu-averages-N6}
are of order unity.

Diagonalizing and ordering the matrix
$\widehat{a}^{\,1}_\text{num-sol}$, we get
new matrices (denoted by a prime), which are
given explicitly in \ref{app:D2-and-N6}.
For the record, we give here the absolute values of the
matrix entries of $\widehat{a}^{\;\prime\,1}_\text{num-sol}$
from \eqref{eq:a1primenumsol-N6-appA}
and the absolute values of the
matrix entries of $\widehat{a}^{\;\prime\,2}_\text{num-sol}$
from \eqref{eq:REa2primenumsol-N6-appA}
and \eqref{eq:IMa2primenumsol-N6-appA},
\bsubeqs\label{eq:a1primenumsol-Absa2primenumsol-N6}
\beqa\label{eq:a1primenumsol-N6}
\hspace*{-9mm}
\text{Abs}\left[\widehat{a}^{\;\prime\,1}_\text{num-sol}\right]
&=&
\text{diag}
\big( 2.03 ,\,  1.69 ,\,  0.653 ,\,  0.569 ,\,  1.21 ,\,  2.59 \big)\,,
\\[2mm]
\label{eq:Absa2primenumsol-N6}
\hspace*{-9mm}
\text{Abs}\left[\widehat{a}^{\;\prime\,2}_\text{num-sol}\right]
&=&
\left(
\renewcommand{\arraycolsep}{0.5pc} 
\renewcommand{\arraystretch}{1.0}  
  \begin{array}{cccccc}
 0.0175 & 1.08 & 0.209 & 0.0608 & 0.0651 & 0.0610 \\
 1.08 & 2.38 & 0.140 & 0.140 & 0.0458 & 0.0541 \\
 0.209 & 0.140 & 1.46 & 0.804 & 0.115 & 0.0341 \\
 0.0608 & 0.140 & 0.804 & 0.459 & 1.25 & 0.201 \\
 0.0651 & 0.0458 & 0.115 & 1.25 & 1.45 & 0.371 \\
 0.0610 & 0.0541 & 0.0341 & 0.201 & 0.371 & 0.965 \\
  \end{array}
\right),
\eeqa
\esubeqs
where we see in the last matrix that the far-off-diagonal elements
(e.g., at positions $15$, $16$, and $26$) are
rather small in comparison to those close to the diagonal,
which is not the case for the original matrix
\eqref{eq:ABSahat2numsol-N6}.
This can be quantified by the following averages and ratio:
\beqa
\label{eq:a1-diag-averages-N6}
\hspace*{-12mm}&&
\Big\{
\langle \text{all} \rangle,\,
\langle \text{band-diag} \rangle,\,
\langle \text{off-band-diag} \rangle,\,
\text{ratio}
\Big\}_{\text{Abs}\left[\widehat{a}^{\;\prime\,2}_\text{num-sol}\right]}^{(N=6)}
=
\nonumber\\[1mm]
\hspace*{-12mm}&&
\left\{ 0.444,\, 0.877,\, 0.0986,\, 8.89 \right\},
\eeqa
where the meaning of the quantities has been explained
on the lines below \eqref{eq:ABSamu-averages-N6}.
The ratio of the band-diagonal value over the off-band-diagonal value
from \eqref{eq:a1-diag-averages-N6} is of order 10,
just as what was found in Sec.~\ref{subsec:D2-and-N4} for $N=4$ .

Similar results are obtained if the other matrix
$\widehat{a}^{\,2}_\text{num-sol}$ is diagonalized and
ordered (the new matrices are denoted by a double prime
and are given explicitly in \ref{app:D2-and-N6}),  
and the corresponding numbers are
\beqa
\label{eq:a2-diag-averages-N6}
\hspace*{-12mm}&&
\Big\{
\langle \text{all} \rangle,\,
\langle \text{band-diag} \rangle,\,
\langle \text{off-band-diag} \rangle,\,
\text{ratio}
\Big\}_{\text{Abs}\left[\widehat{a}^{\;\prime\prime\,1}_\text{num-sol}\right]}^{(N=6)}
=
\nonumber\\[1mm]
\hspace*{-12mm}&&
\left\{ 0.411,\, 0.774,\, 0.121,\, 6.40 \right\}\,.
\eeqa
For both choices of the bases,
we see a more or less band-diagonal structure
appearing if one of the matrices is diagonalized.

In closing, we comment briefly on the results  
$\widehat{a}^{\,1,\,2}_\text{num-sol}$ for $D=2$ and $N=6$.
The entries of these matrices (70 real numbers in total)
were obtained by the numerical minimization routine \texttt{FindMinimum} 
of \textsc{Mathematica} 12.1 (cf. Ref.~\refcite{Wolfram1991}).
This minimization operates on an auxiliary function,
which consists of a sum of 70 squares, each
square containing the real or imaginary part of one of
the components of the simplified algebraic matrix equation
\eqref{eq:simplified-equation}.
(The auxiliary function has a size of about $7$ MB,
as simplifications are difficult to obtain.)
The accuracy of the obtained 70 numbers
can, in principle, be increased arbitrarily.
Hence, given the exact (pseudorandom)
constants \eqref{eq:phatmunum-etahatmunumsol-N6-appA}, the
obtained matrices \eqref{eq:ahatmunumsol-N6-appA}
may be called ``quasi-exact.''


\subsection{Corresponding bosonic master-field matrices}
\label{subsec:Corresponding-bosonic-master-field-matrices}

Up till now, we have focused on the
bosonic matrices
$\widehat{a}^{\;\rho}$.
The corresponding master-field matrices
$\widehat{A}^{\;\rho}$ follow from
\eqref{eq:IIB-matrix-model-master-field}.
That expression can be rewritten as follows:
\bsubeqs\label{eq:Ahatrho-with-D}
\beqa
\widehat{A}^{\;\rho}
&=&
D \cdot \widehat{a}^{\;\rho} \cdot D^{-1}\,,
\\[2mm]
D_{kl} &\equiv&
\Big[ \text{diag}
\left(
e^{\displaystyle{i\,\widehat{p}_{1} \tau_\text{eq}}},\,
\ldots \,,\,
e^{\displaystyle{i\,\widehat{p}_{N} \tau_\text{eq}}}
\right)
\Big]_{kl}\,,
\eeqa
\esubeqs
where we suppress the dependence of $\tau_\text{eq}$
in $D$ and  $\widehat{A}^{\;\rho}$.

Explicit matrices $\widehat{a}^{\,1,\,2}_\text{num-sol}$
were obtained in Secs.~\ref{subsec:D2-and-N4} 
and \ref{subsec:D2-and-N6}from the simplified algebraic equation  
\eqref{eq:simplified-equation-gtildemunu-stilde-Eucl} for $D=2$ and 
with a particular realization of the pseudorandom
constants $\widehat{p}$ and $\widehat{\eta}^{\;\rho}$.
The corresponding master-field matrices are
\bsubeqs\label{eq:Ahatrhonum-with-Dnum}
\beqa
\label{eq:Ahatrhonum}
\widehat{A}^{\;\rho}_\text{num-sol}
&=&
D_\text{num} \cdot \widehat{a}_\text{num-sol}^{\;\rho} \cdot D_\text{num}^{-1}\,,
\\[2mm]
\label{eq:Dnum}
\Big[D_\text{num} \Big]_{kl} &\equiv&
\Big[ \text{diag}
\left(
e^{\displaystyle{i\,\widehat{p}_{1,\,\text{num}}\,\tau_\text{eq,\,num}}},\,
  \ldots \,,\,
e^{\displaystyle{i\,\widehat{p}_{N,\,\text{num}}\,\tau_\text{eq,\,num}}}
\right)
\Big]_{kl}\,,
\eeqa
\esubeqs
where $\tau_\text{eq,\,num}$ is an appropriate numerical value
for $\tau_\text{eq}$.
[We conjecture that the value of $\tau_\text{eq,\,num}$
must be so large that the diagonal entries in \eqref{eq:Dnum},
for given values of $\widehat{p}_{k,\,\text{num}}$,
cover the unit circle in the complex plane more or less uniformly
in the limit $N\to\infty$.]

With the similarity transformation \eqref{eq:ahatrho-similarity-transformed} 
on $\widehat{a}^{\;\rho}_\text{num-sol}$ to diagonalize the
$\rho=1$ matrix (which requires $S=S_{1}$), we get
from \eqref{eq:Ahatrhonum-with-Dnum}
\bsubeqs\label{eq:Ahatrhonum-diag-structure-T1}
\beqa
\label{eq:Ahatrhonum-diag-structure}
\widehat{A}_\text{num-sol}^{\;\prime\,\rho}
&=&
T_{1} \cdot \widehat{a}^{\;\prime\,\rho}_\text{num-sol} \cdot T_{1}^{-1}\,,
\\[2mm]
\label{eq:T1}
T_{1} &\equiv&
D_\text{num} \cdot S_{1}^{-1}\,,
\eeqa
\esubeqs
and similarly for the other diagonalization
$\widehat{a}^{\;\prime\,\prime\,\rho}_\text{num-sol}$
from $S=S_{2}$.

From the expression \eqref{eq:Ahatrhonum-diag-structure},
we conclude that the diagonal/band-diagonal
structure discovered for $\widehat{a}^{\;\prime\,\rho}_\text{num-sol}$
in \eqref{eq:amuprimenumsol-N4} and \eqref{eq:amuprimenumsol-N6-appA}
directly carries over to the  master-field
matrices $\widehat{A}_\text{num-sol}^{\;\prime\,\rho}\,$.
But we prefer to focus the discussion of 
this  
paper on the $\tau$-independent matrices $\widehat{a}^{\;\rho}$,
as a proper value for $\tau_\text{eq,\,num}$ is not required
for that discussion.


\section{Conclusion}
\label{sec:Conclusion}

The large-$N$ bosonic master field \eqref{eq:IIB-matrix-model-master-field}
for bosonic observables in the IIB matrix model
is essentially determined as a solution of
the algebraic equation \eqref{eq:IIB-matrix-model-algebraic-equation}.
To solve that equation is a formidable task and we have,
instead, considered the simplified algebraic
equation \eqref{eq:simplified-equation-gtildemunu-stilde-Eucl}.

For low dimensionality ($D=2$) and small matrix sizes ($N=4$ and $6$),
we have established the existence of one or more nontrivial solutions
of that simplified algebraic equation
for an explicit realization of the pseudorandom constants;
see Secs.~\ref{subsec:D2-and-N4} and \ref{subsec:D2-and-N6}.
The obtained matrices do not show an obvious band-diagonal structure.
But if one of these two matrices is diagonalized (with ordered
eigenvalues), then the other matrix does get a band-diagonal structure.
For $N$=6, this has been quantified by the averages and ratios given
in \eqref{eq:a1-diag-averages-N6} and \eqref{eq:a2-diag-averages-N6}.

There are, however, many open questions and let us mention two.
First, how does the appropriately defined
width $\Delta N$ of the band diagonal in $\widehat{a}^{\;\prime\,2}$
(or in $\widehat{a}^{\;\prime\prime\,1}$) depend
on $N$ and how do the off-band-diagonal entries scale with $N$?
Second, is there indeed no essential difference for the
appearance of a diagonal/band-diagonal structure between
the Euclidean ($\widetilde{s}=1$) and
the Lorentzian ($\widetilde{s}=-1$) models?
For the first question, we need to consider larger and larger
values of $N$. For the second question, we need to go to larger
dimensionality, for example, $D=4$ or $10$.
Insight into both extensions (larger $N$ and larger $D$)
may perhaps come from approximative results;
cf. the earlier work reported in Ref.~\refcite{AlbertyGreensite1984}.

Also missing is a convincing explanation for why the matrix
solution of the simplified algebraic equation
\eqref{eq:simplified-equation-gtildemunu-stilde-Eucl}
has a hidden diagonal/band-diagonal structure
(some preliminary considerations are presented
in \ref{app:Possible-large-N-behavior}).
Still, the diagonal/band-diagonal structure found in the
``quasi-exact'' solutions from Secs.~\ref{subsec:D2-and-N4}
and \ref{subsec:D2-and-N6}, admittedly only for low dimensionality
and relatively small values of the matrix size,
provides clear evidence for one of the assumption in
our previous discussion of spacetime extraction
from the bosonic IIB-matrix-model master field.

\section*{Acknowledgments}

It is a pleasure to thank J. Nishimura
for informative discussions over the last years
and the referee of the present paper for constructive remarks.

\section*{Note Added in Proof}

We have been informed by T. Fischbacher\cite{Fischbacher2021B}
that he and his colleagues at Google \mbox{Research,} Z\"{u}rich,
have obtained numerical solutions of the algebraic
equation \eqref{eq:simplified-equation-gtildemunu-stilde-Eucl}
for $(D,\,N)=(10,\,50)$ and that these solutions display
a diagonal/band-diagonal structure.

\begin{appendix}

\section{Matrices for $D=2$ and $N=6$}
\label{app:D2-and-N6}


The matrices from Sec.~\ref{subsec:D2-and-N6} are rather big
and it is better to show their real and imaginary parts separately.

Taking the following pseudorandom constants:
\bsubeqs\label{eq:phatmunum-etahatmunumsol-N6-appA}
\beqa
\hspace*{-11mm}&&
\widehat{p}_\text{num}
=
\left\{ \frac{53}{500},\;-\frac{9}{100},\;-\frac{441}{1000},\;
        \frac{217}{1000},\;\frac{371}{1000},\; \frac{19}{40} \right\}\,,
\eeqa
\beqa
\hspace*{-11mm}&&
\text{Re}\left[\widehat{\eta}^{\,1}_\text{num}\right]
=\left(   
\renewcommand{\arraycolsep}{0.5pc} 
\renewcommand{\arraystretch}{2.0}  
\begin{array}{cccccc}
 -\fraclarge{81}{125} & \fraclarge{71}{1000} & -\fraclarge{151}{500} & \fraclarge{371}{500} & -\fraclarge{83}{200} & \fraclarge{491}{1000} \\
 \fraclarge{71}{1000} & -\fraclarge{279}{1000} & -\fraclarge{259}{500} & -\fraclarge{13}{1000} & -\fraclarge{493}{500} & \fraclarge{449}{1000} \\
 -\fraclarge{151}{500} & -\fraclarge{259}{500} & -\fraclarge{413}{1000} & \fraclarge{911}{1000} & \fraclarge{203}{250} & \fraclarge{299}{1000} \\
 \fraclarge{371}{500} & -\fraclarge{13}{1000} & \fraclarge{911}{1000} & \fraclarge{671}{1000} & -\fraclarge{417}{500} & -\fraclarge{913}{1000} \\
 -\fraclarge{83}{200} & -\fraclarge{493}{500} & \fraclarge{203}{250} & -\fraclarge{417}{500} & \fraclarge{51}{125} & \fraclarge{181}{250} \\
 \fraclarge{491}{1000} & \fraclarge{449}{1000} & \fraclarge{299}{1000} & -\fraclarge{913}{1000} & \fraclarge{181}{250} & \fraclarge{261}{1000} \\
\end{array}
\right),
\eeqa
\beqa
\hspace*{-11mm}&&
\text{Im}\left[\widehat{\eta}^{\,1}_\text{num}\right]
=
\left(
\renewcommand{\arraycolsep}{0.5pc} 
\renewcommand{\arraystretch}{2.0}  
\begin{array}{cccccc}
 0 & -\fraclarge{441}{1000} & -\fraclarge{17}{250} & -\fraclarge{87}{1000} & -\fraclarge{127}{200} & -\fraclarge{199}{500} \\
 \fraclarge{441}{1000} & 0 & -\fraclarge{177}{250} & -\fraclarge{783}{1000} & -\fraclarge{303}{500} & \fraclarge{969}{1000} \\
 \fraclarge{17}{250} & \fraclarge{177}{250} & 0 & -\fraclarge{14}{25} & \fraclarge{259}{1000} & -\fraclarge{711}{1000} \\
 \fraclarge{87}{1000} & \fraclarge{783}{1000} & \fraclarge{14}{25} & 0 & \fraclarge{43}{250} & \fraclarge{1}{125} \\
 \fraclarge{127}{200} & \fraclarge{303}{500} & -\fraclarge{259}{1000} & -\fraclarge{43}{250} & 0 & -\fraclarge{491}{1000} \\
 \fraclarge{199}{500} & -\fraclarge{969}{1000} & \fraclarge{711}{1000} & -\fraclarge{1}{125} & \fraclarge{491}{1000} & 0 \\
\end{array}
\right),
\eeqa
\beqa
\hspace*{-11mm}&&
\text{Re}\left[\widehat{\eta}^{\,2}_\text{num}\right]
=
\left(
\renewcommand{\arraycolsep}{0.5pc} 
\renewcommand{\arraystretch}{2.0}  
\begin{array}{cccccc}
 \fraclarge{41}{200} & \fraclarge{53}{1000} & -\fraclarge{241}{250} & \fraclarge{621}{1000} & \fraclarge{3}{20} & -\fraclarge{51}{200} \\
 \fraclarge{53}{1000} & -\fraclarge{139}{500} & \fraclarge{23}{200} & -\fraclarge{557}{1000} & \fraclarge{7}{100} & -\fraclarge{137}{200} \\
 -\fraclarge{241}{250} & \fraclarge{23}{200} & -\fraclarge{31}{500} & -\fraclarge{22}{125} & \fraclarge{14}{25} & -\fraclarge{31}{100} \\
 \fraclarge{621}{1000} & -\fraclarge{557}{1000} & -\fraclarge{22}{125} & \fraclarge{289}{1000} & -\fraclarge{227}{1000} & -\fraclarge{103}{200} \\
 \fraclarge{3}{20} & \fraclarge{7}{100} & \fraclarge{14}{25} & -\fraclarge{227}{1000} & \fraclarge{17}{1000} & \fraclarge{369}{1000} \\
 -\fraclarge{51}{200} & -\fraclarge{137}{200} & -\fraclarge{31}{100} & -\fraclarge{103}{200} & \fraclarge{369}{1000} & -\fraclarge{171}{1000} \\
\end{array}
\right),
\eeqa
\beqa
\hspace*{-11mm}&&
\text{Im}\left[\widehat{\eta}^{\,2}_\text{num}\right]
=
\left(
\renewcommand{\arraycolsep}{0.5pc} 
\renewcommand{\arraystretch}{2.0}  
\begin{array}{cccccc}
 0 & \fraclarge{449}{500} & -\fraclarge{31}{250} & \fraclarge{233}{1000} & -\fraclarge{413}{500} & -\fraclarge{807}{1000} \\
 -\fraclarge{449}{500} & 0 & -\fraclarge{56}{125} & \fraclarge{7}{50} & -\fraclarge{77}{200} & \fraclarge{23}{500} \\
 \fraclarge{31}{250} & \fraclarge{56}{125} & 0 & \fraclarge{409}{500} & \fraclarge{57}{250} & -\fraclarge{689}{1000} \\
 -\fraclarge{233}{1000} & -\fraclarge{7}{50} & -\fraclarge{409}{500} & 0 & \fraclarge{189}{500} & -\fraclarge{953}{1000} \\
 \fraclarge{413}{500} & \fraclarge{77}{200} & -\fraclarge{57}{250} & -\fraclarge{189}{500} & 0 & \fraclarge{47}{200} \\
 \fraclarge{807}{1000} & -\fraclarge{23}{500} & \fraclarge{689}{1000} & \fraclarge{953}{1000} & -\fraclarge{47}{200} & 0 \\
\end{array}
\right),
\eeqa
\esubeqs
we obtain from the simplified algebraic
equation \eqref{eq:simplified-equation-gtildemunu-stilde-Eucl},
for $D=2$ and $N=6$, the following numerical solution:
\bsubeqs\label{eq:ahatmunumsol-N6-appA}
\beqa
\hspace*{-9mm}&&
\text{Re}\left[\widehat{a}^{\,1}_\text{num-sol}\right]
=
\left(
\begin{array}{cccccc}
 0.234 & 0.346 & 0.578 & 0.328 & 0.336 & -0.152 \\
 0.346 & -0.721 & -0.698 & 0.0277 & 0.483 & -0.344 \\
 0.578 & -0.698 & 0.445 & -0.208 & -0.466 & 0.613 \\
 0.328 & 0.0277 & -0.208 & 0.0504 & 0.0989 & 0.543 \\
 0.336 & 0.483 & -0.466 & 0.0989 & -0.256 & -0.319 \\
 -0.152 & -0.344 & 0.613 & 0.543 & -0.319 & 0.248 \\
\end{array}
\right)\,,
\eeqa
\beqa
\hspace*{-9mm}&&
\text{Im}\left[\widehat{a}^{\,1}_\text{num-sol}\right]
=
\left(
\begin{array}{cccccc}
 0 & 0.231 & 0.267 & 1.08 & 0.760 & -0.176 \\
 -0.231 & 0 & 0.643 & -0.566 & -0.374 & -0.125 \\
 -0.267 & -0.643 & 0 & 0.724 & 1.11 & 0.108 \\
 -1.08 & 0.566 & -0.724 & 0 & -0.0565 & -0.301 \\
 -0.760 & 0.374 & -1.11 & 0.0565 & 0 & 0.545 \\
 0.176 & 0.125 & -0.108 & 0.301 & -0.545 & 0 \\
\end{array}
\right)\,,
\eeqa
\beqa
\hspace*{-9mm}&&\text{Re}\left[\widehat{a}^{\,2}_\text{num-sol}\right]
=
\left(
\begin{array}{cccccc}
 -0.560 & 0.820 & 0.493 & -0.348 & 0.377 & -0.126 \\
 0.820 & -0.866 & -0.430 & -0.714 & 0.489 & 1.10 \\
 0.493 & -0.430 & -0.623 & -0.0737 & 0.0523 & 0.451 \\
 -0.348 & -0.714 & -0.0737 & 0.459 & -0.279 & 0.763 \\
 0.377 & 0.489 & 0.0523 & -0.279 & 0.400 & -0.195 \\
 -0.126 & 1.10 & 0.451 & 0.763 & -0.195 & 1.19 \\
\end{array}
\right)\,,
\eeqa
\beqa
\hspace*{-9mm}&&\text{Im}\left[\widehat{a}^{\,2}_\text{num-sol}\right]
=
\left(
\begin{array}{cccccc}
 0 & 0.429 & 0.266 & 0.198 & -0.421 & 0.607 \\
 -0.429 & 0 & 0.554 & -0.388 & 0.388 & -0.139 \\
 -0.266 & -0.554 & 0 & -1.04 & -0.0679 & -0.431 \\
 -0.198 & 0.388 & 1.04 & 0 & -0.378 & 0.498 \\
 0.421 & -0.388 & 0.0679 & 0.378 & 0 & 0.444 \\
 -0.607 & 0.139 & 0.431 & -0.498 & -0.444 & 0 \\
\end{array}
\right)
\,,
\eeqa
\esubeqs
where only three  
significant digits are shown.

Diagonalizing and ordering the matrix
$\widehat{a}^{\,1}_\text{num-sol}$ gives
(the new matrices are denoted by a prime)
\bsubeqs\label{eq:amuprimenumsol-N6-appA}
\beqa\label{eq:a1primenumsol-N6-appA}
\hspace*{-10mm}&&
\widehat{a}^{\;\prime\,1}_\text{num-sol}
=
\text{diag}\big(-2.03,\, -1.69,\, -0.653,\, 0.569,\, 1.21,\, 2.59 \big)\,,
\eeqa
\beqa
\label{eq:REa2primenumsol-N6-appA}
\hspace*{-10mm}&&
\text{Re}\left[\widehat{a}^{\;\prime\,2}_\text{num-sol}\right]
=
\left(
\begin{array}{cccccc}
 -0.0175 & 0.491 & -0.195 & 0.0240 & -0.0384 & -0.0596 \\
 0.491 & -2.38 & -0.140 & 0.108 & -0.0356 & -0.0540 \\
 -0.195 & -0.140 & 1.46 & 0.642 & 0.0403 & 0.00702 \\
 0.0240 & 0.108 & 0.642 & 0.459 & 1.10 & -0.0952 \\
 -0.0384 & -0.0356 & 0.0403 & 1.10 & 1.45 & -0.229 \\
 -0.0596 & -0.0540 & 0.00702 & -0.0952 & -0.229 & -0.965 \\
\end{array}
\right),
\eeqa
\beqa
\label{eq:IMa2primenumsol-N6-appA}
\hspace*{-10mm}&&
\text{Im}\left[\widehat{a}^{\;\prime\,2}_\text{num-sol}\right]
=
\left(\!
\begin{array}{cccccc}
 0 & -0.962 & -0.0748 & -0.0559 & 0.0525 & 0.0129 \\
 0.962 & 0 & -0.00653 & 0.0886 & -0.0287 & 0.00331 \\
 0.0748 & 0.00653 & 0 & -0.485 & -0.108 & -0.0334 \\
 0.0559 & -0.0886 & 0.485 & 0 & 0.604 & 0.177 \\
 -0.0525 & 0.0287 & 0.108 & -0.604 & 0 & -0.293 \\
 -0.0129 & -0.00331 & 0.0334 & -0.177 & 0.293 & 0 \\
\end{array}
\!\right)\!,
\eeqa
\esubeqs
where, again, only three  
significant digits are shown.
Similarly, diagonalizing and ordering the matrix
$\widehat{a}^{\,2}_\text{num-sol}$ gives
(the new matrices are denoted by a double prime)
\bsubeqs\label{eq:amuprimeprimenumsol-N6-appA}
\beqa
\hspace*{-12mm}&&
\text{Re}\left[\widehat{a}^{\;\prime\,\prime\,1}_\text{num-sol}\right]
=
\left(
\begin{array}{cccccc}
 -1.73 & -0.00903 & 0.115 & 0.0429 & 0.0344 & -0.0108 \\
 -0.00903 & 2.08 & 0.736 & -0.151 & 0.0896 & -0.00741 \\
 0.115 & 0.736 & 0.967 & 0.00290 & 0.265 & -0.0423 \\
 0.0429 & -0.151 & 0.00290 & -1.81 & 0.256 & -0.0382 \\
 0.0344 & 0.0896 & 0.265 & 0.256 & -0.111 & -0.385 \\
 -0.0108 & -0.00741 & -0.0423 & -0.0382 & -0.385 & 0.608 \\
\end{array}
\right),
\eeqa
\beqa
\hspace*{-12mm}&&
\text{Im}\left[\widehat{a}^{\;\prime\,\prime\,1}_\text{num-sol}\right]
=
\left(
\begin{array}{cccccc}
 0 & 0.0384 & -0.0693 & 0.0689 & -0.0424 & -0.0215 \\
 -0.0384 & 0 & -0.463 & 0.135 & 0.108 & 0.00611 \\
 0.0693 & 0.463 & 0 & -0.579 & -0.337 & 0.0669 \\
 -0.0689 & -0.135 & 0.579 & 0 & 0.211 & -0.0408 \\
 0.0424 & -0.108 & 0.337 & -0.211 & 0 & 0.606 \\
 0.0215 & -0.00611 & -0.0669 & 0.0408 & -0.606 & 0 \\
\end{array}
\right),
\eeqa
\beqa
\hspace*{-12mm}&&
\widehat{a}^{\;\prime\,\prime\,2}_\text{num-sol}
=
\text{diag}\big( -2.81,\,  -1.19,\, -0.429,\, 0.413,\, 1.44,\, 2.58\big)\,,
\eeqa
\esubeqs
with three  
significant digits shown.
Further discussion of these results appears in Sec.~\ref{subsec:D2-and-N6}.


\section{Possible Large-$N$ Behavior}
\label{app:Possible-large-N-behavior}


We have found, in Sec.~\ref{sec:Numerical solutions},
a clear hint of a diagonal/band-diagonal structure in the solutions
of the simplified algebraic equation
\eqref{eq:simplified-equation-gtildemunu-stilde-Eucl},
albeit at low dimensionality ($D=2$) and small
matrix size ($N=4$ or $6$). Even while keeping
$D=2$, we do not really know if this structure
survives in the large-$N$ limit. In this appendix, we
present some preliminary ideas.

The simplified algebraic equation
\eqref{eq:simplified-equation-gtildemunu-stilde-Eucl},
for fixed values of $\widehat{p}_{k}$
and $\widehat{\eta}^{\;\rho}_{\;kl}$,
is surprisingly difficult to solve and understand.
Some minor progress can be made if we consider a
judicial approximation.

Take the absolute values of the master-noise matrix entries
$\widehat{\eta}^{\;\rho}_{\;kl}$ to be of order unity
and consider the absolute values of the
master momenta $\widehat{p}_{k}$
to be very much smaller than unity [in principle, these
small values can be compensated by considering very
much larger values of $\tau_\text{eq}$ in \eqref{eq:Ahatrho-with-D}].
Then, it makes sense to look at
the following \emph{approximate} algebraic equation:
\beq
\label{eq:approx-alg-eq-appB}
0=
\Big[\widehat{a}_{\,\nu},\,\big[\widehat{a}^{\,\nu},\,
\widehat{a}^{\;\rho}\big]\Big]+\widehat{\eta}^{\;\rho}\,,
\eeq
where the matrix indices have been suppressed altogether
and where an index $\nu$ has been lowered with
the Euclidean metric $\widetilde{g}_{\mu\nu}$ from
\eqref{eq:simplified-equation-gtildemunu} and
\eqref{eq:simplified-equation-stilde-Eucl}.

If we make a similarity transformation \eqref{eq:ahatrho-similarity-transformed}
with a matrix $S=\widetilde{S}_{1}$,
then \eqref{eq:approx-alg-eq-appB} keeps the same
form if the master-noise matrix is transformed accordingly,
\bsubeqs\label{eq:approx-alg-eq-prime-appB}
\beqa
0&=&
\Big[\widehat{a}^{\;\prime}_{\,\nu},\,
\big[\widehat{a}^{\;\prime\,\nu},\,
\widehat{a}^{\;\prime\,\rho}\big]\Big]
+\widehat{\eta}^{\;\prime\,\rho}\,,
\\[2mm]
\widehat{\eta}^{\;\prime\,\rho}
&=& \widetilde{S}_{1} \cdot \widehat{\eta}^{\;\rho} \cdot \widetilde{S}_{1}^{-1}\,,
\eeqa
\esubeqs
where $\widehat{a}^{\;\prime\,1}$ is now assumed to be diagonal,
hence the notation with the single prime.

With given master-noise matrices $\widehat{\eta}^{\;\rho}$,
the goal is to solve equation \eqref{eq:approx-alg-eq-appB}
 for the unknown matrices $\widehat{a}^{\;\rho}$.
Let us also assume that we have
\beq
\label{eq:D2-appB}
D=2\,,
\eeq
so that there are only two unknown matrices in the problem.

This problem is still difficult. So, let us, instead,
consider the \emph{inverse} problem: make certain
\emph{Ans\"{a}tze} $\widehat{a}^{\;1}_\text{Ansatz}$ and
$\widehat{a}^{\;2}_\text{Ansatz}$,
calculate the corresponding
matrices $\widehat{\eta}^{\;\rho}_\text{result}$
from \eqref{eq:approx-alg-eq-appB},
and then ask if these calculated matrices are
more or less noise-like (all entries pseudorandom and with an
absolute value of order unity).

In fact, we will consider the approximate algebraic
equation \eqref{eq:approx-alg-eq-prime-appB}
with diagonal $\widehat{a}^{\;\prime\,1}$.
Making appropriate \emph{Ans\"{a}tze}
$\widehat{a}^{\;\prime\,1}_\text{Ansatz}$ and
$\widehat{a}^{\;\prime\,2}_\text{Ansatz}$,
we calculate the resulting matrices
\beq
\label{eq:etahatprimerho-result-appB}
\widehat{\eta}^{\;\prime\,\rho}_\text{result}
= - \Big[\widehat{a}^{\;\prime}_{\,\nu\;\text{Ansatz}}\,,\,
\big[\widehat{a}^{\;\prime\,\nu}_\text{Ansatz}\,,\,
\widehat{a}^{\;\prime\,\rho}_\text{Ansatz}\big]\Big]\,.
\eeq
For $\widehat{a}^{\;\prime\,1}_\text{Ansatz}$,
we take a diagonal matrix with ordered eigenvalues
and, for $\widehat{a}^{\;\prime\,2}_\text{Ansatz}$,
we take the sum of a real band-diagonal matrix
(with width $\Delta N=3$, in order to be specific) and a purely
random matrix in the bulk.
Both of these \emph{Ansatz} matrices are traceless
and Hermitian. We will try to scale the entries of the matrices
$\widehat{a}^{\;\prime\,\rho}_\text{Ansatz}$
by appropriate powers of $N$,
so that the entries of the resulting matrices
\eqref{eq:etahatprimerho-result-appB} are more or less
random and have average absolute values
which are approximately constant as $N$ becomes very large.

Specifically, we use the following construction for the
matrices $\widehat{a}^{\;\prime\,\rho}_\text{Ansatz}$.
Assume, for simplicity, $N$ to be even.
With the matrix indices $k,\,l$
running over $\{  1,\, \, \ldots \, N \}$
and the directional index $\rho$
running over $\{  1,\,  \ldots \,,\,D \}$ for $D=2$,
we then define
\bsubeqs\label{eq:a-prime-rho-Ansatz-construction-appB}
\beqa\label{eq:a-prime-1-tmp-appB}
\hspace*{-12mm}&&
\Big[\widehat{a}^{\;\prime\,1}_\text{tmp}\Big]_{kl} =
\Xi^{2}\,N^{-2/3}\;    
\Big[ \text{diag}
\big(  -N/2,\,
\ldots \,, -2,\, -1,\,  1,\, 2,\, \ldots \, ,\,
\,N/2  \big)
\Big]_{kl}\,,
\\[2mm]
\label{eq:a-prime-2-tmp-appB}
\hspace*{-12mm}&&
\Big[\widehat{a}^{\;\prime\,2}_\text{tmp}\Big]_{kl} =
\Big[\widehat{a}^{\;\prime\,2}_\text{tmp-rand}\Big]_{kl}
+
\Big[\widehat{a}^{\;\prime\,2}_\text{tmp-band}\Big]_{kl}\,,
\\[2mm]
\hspace*{-12mm}&&
\Big[\widehat{a}^{\;\prime\,2}_\text{tmp-rand}\Big]_{kl}
=
\nonumber\\[1mm]
\hspace*{-12mm}&&
\Xi^{-1}\,N^{-1/3}\;   
\;\Big(\text{randominteger}[-1,\,+1]+i\,\text{randominteger}[-1,\,+1]\Big)
\,,
\\[2mm]
\hspace*{-12mm}&&
\Big[\widehat{a}^{\;\prime\,2}_\text{tmp-band}\Big]_{kl}
=
\nonumber\\[1mm]
\hspace*{-12mm}&&
\begin{cases}
\Xi^{-4} \,N^{-2/3}   
\;\text{randominteger}[-N/2,\,+N/2] \,,
&   \text{for}\;\; k>l+1  \;\; \wedge \;\; l<k+1 \,,
 \\[2mm]
0 \,,   &  \text{otherwise} \,,
\end{cases}
\\[2mm]
\hspace*{-12mm}&&
N = 2\,K\,, \;\;\text{for}\;\;K \in \mathbb{N}^{+}\,,
\\[2mm]
\hspace*{-12mm}&&
\Xi \equiv 126/100\,,
\eeqa
and make all matrices traceless ($\widehat{a}^{\;\prime\,1}_\text{tmp}$
is already traceless),
\beq
\label{eq:a-prime-rho-traceless-appB}
\widehat{a}^{\;\prime\,\rho}_\text{tmp-traceless}
=
\widehat{a}^{\;\prime\,\rho}_\text{tmp}
- \frac{1}{N}\;\text{Tr} \left(\widehat{a}^{\;\prime\,\rho}_\text{tmp}\right)\;
\id_{N}\,,
\eeq
and Hermitian,
\beq
\label{eq:a-prime-rho-Ansatz-appB-appB}
\widehat{a}^{\;\prime\,\rho}_\text{Ansatz}
=
\frac{1}{2}\;\left[\widehat{a}^{\;\prime\,\rho}_\text{tmp-traceless}
+\left(\widehat{a}^{\;\prime\,\rho}_\text{tmp-traceless}\right)^{\dagger}\right]
\,.
\eeq
\esubeqs  

Observe that the diagonal entries of
$\widehat{a}^{\;\prime\,1}_\text{Ansatz}$
and the band-diagonal entries of
$\widehat{a}^{\;\prime\,2}_\text{Ansatz}$
grow as $\text{O}\left(N^{1/3}\right)$,
while the off-band-diagonal entries of
$\widehat{a}^{\;\prime\,\rho}_\text{Ansatz}$
drop as $\text{O}\left(N^{-1/3}\right)$.
Incidentally, the \textit{Ansatz}
\eqref{eq:a-prime-rho-Ansatz-construction-appB} is strictly rational
if we take $N=2^{3\,p}$, for positive integer $p$.

\begin{table}[t]
\tbl{Matrices $\widehat{\eta}^{\;\prime\,\rho}_\text{result}$
calculated from \eqref{eq:etahatprimerho-result-appB}
with \textit{Ans\"{a}tze} \eqref{eq:a-prime-rho-Ansatz-construction-appB}.
Shown are representative results for  
the minimal value, the maximal value, and the mean value
of the absolute values of the matrix entries (the matrix size is $N$).
\vspace*{2mm}}
{\renewcommand{\tabcolsep}{1.25pc}    
\renewcommand{\arraystretch}{1.5}   
\begin{tabular}{l|c|c}
\hline\hline
  &  $\text{Abs}\left[\widehat{\eta}^{\;\prime\,1}_\text{result}\right]$
  &  $\text{Abs}\left[\widehat{\eta}^{\;\prime\,2}_\text{result}\right]$  \\
\hline
  &  \{min , max , mean\} & \{min , max , mean\} \\
\hline\hline
$N$ = 8   & \{0.158 , 3.92 , 1.14\}      & \{0 , 4.33 , 0.851\}\\
%
%
$N$ = 64  & \{0.0227 ,  12.1 , 1.04\}    & \{0 , 9.62 , 1.05\}\\
$N$ = 512 & \{0.00206 , 45.4 , 1.38\}    & \{0 , 21.8 , 1.94\}\\
\hline\hline
\end{tabular}
\label{tab-etahatprime-result-num}
}
\end{table}

With the \textit{Ansatz} matrices $\widehat{a}^{\;\prime\,\rho}_\text{Ansatz}$
from \eqref{eq:a-prime-rho-Ansatz-construction-appB},
we calculate the matrices $\widehat{\eta}^{\;\prime\,\rho}_\text{result}$
from \eqref{eq:etahatprimerho-result-appB}.
For $N=8$, $64$, and $512$,
we have some representative results which show that the matrices
$\widehat{\eta}^{\;\prime\,\rho}_\text{result}$ are more or less
noise-like, except that
$\widehat{\eta}^{\;\prime\,1}_\text{result}$ has somewhat large entries
on the diagonal and that
$\widehat{\eta}^{\;\prime\,2}_\text{result}$
has zeros on the diagonal and relatively small values in a wide band
around the diagonal.
Still, the matrices relevant to the original
problem \eqref{eq:approx-alg-eq-appB} are obtained
by a similarity transformation,
\beq
\label{eq:etahat-result-appB}
\widehat{\eta}^{\;\rho}_\text{result}
= \widetilde{S}_{1}^{-1} \cdot \widehat{\eta}^{\;\prime\,\rho}_\text{result}
  \cdot \widetilde{S}_{1}\,,
\eeq
which will change the diagonal values of, in particular,
$\widehat{\eta}^{\;\prime\,2}_\text{result}$.
More important is that the calculated matrices
$\widehat{\eta}^{\;\prime\,\rho}_\text{result}$
have entries that do not seem to grow drastically with $N$.
Indeed, the arithmetic mean (average)
of the absolute values of the entries of the calculated matrices
$\widehat{\eta}^{\;\prime\,\rho}_\text{result}$
stays more or less constant at unity
for values of $N$ up to $512$;
see Table~\ref{tab-etahatprime-result-num}.
Admittedly, the results from Table~\ref{tab-etahatprime-result-num}
are not perfect (the maximum values grow  with $N$,
as do the mean values  to a lesser extent), but perhaps the
\textit{Ansatz} can be improved, as will be discussed in the next
paragraph.

Based on the results obtained above, we conjecture that
the solution matrices $\widehat{a}^{\;\rho}$ of the algebraic
equation \eqref{eq:simplified-equation-gtildemunu-stilde-Eucl}
may display, after diagonalization of one of them,
a  diagonal/band-diagonal structure, having
off-band-diagonal entries that drop in magnitude with $N$
(perhaps as $N^{-1/3}$)
and diagonal entries that grow with $N$
(perhaps as $N^{1/3}$).
Here, the assumption is that the 
left-hand side  
of \eqref{eq:simplified-equation-gtildemunu-stilde-Eucl} can,
in first approximation, be neglected.
Two of our further assumptions are perhaps less adequate.
First, the assumption of a constant band-diagonal width $\Delta N=3$
in $\widehat{a}^{\;\prime\,2}_\text{Ansatz}$ may be invalid and
perhaps $\Delta N$ increases with a small positive power
of $N$.
Second, the assumption of a constant range for the entries on the
band diagonal in $\widehat{a}^{\;\prime\,2}$
may also be invalid and an improvement would
have an over-all structure matching the
ordered eigenvalues of $\widehat{a}^{\;\prime\,1}$.
Still, it appears not at all impossible that some type of
diagonal/band-diagonal structure in the matrices
$\widehat{a}^{\;\prime\,\rho}$ remains in the large-$N$ limit.

\end{appendix}


\end{document}